\documentstyle[draft]{mn}

\def\simgt{\mathrel{\raise0.35ex\hbox{$\scriptstyle >$}\kern-0.6em
\lower0.40ex\hbox{{$\scriptstyle \sim$}}}}
\def\simlt{\mathrel{\raise0.35ex\hbox{$\scriptstyle <$}\kern-0.6em
\lower0.40ex\hbox{{$\scriptstyle \sim$}}}}

\input epsf

\begin{document}
 
\title[Difficulties with MOND]{Cosmological Difficulties with Modified
Newtonian Dynamics (or: La Fin du MOND?)}

\author[D.~Scott, M.~White, J.~Cohn, \& E.~Pierpaoli]
{Douglas Scott${}^1$, Martin White${}^2$, Joanne D.~Cohn${}^2$,
and Elena Pierpaoli${}^1$\\
${}^1$Department of Physics and Astronomy, University of British Columbia,
BC V6T1Z1~~Canada\\
${}^2$Harvard-Smithsonian Center for Astrophysics, Cambridge, MA 02138}

\date{Accepted ... ;
      Received ... ;
      in original form ...}

\pagerange{000--000}

\maketitle

\begin{abstract}
The cold dark matter paradigm has been extremely successful for explaining
a wide range of cosmological phenomena.  Nevertheless, since evidence for
non-baryonic dark matter remains indirect, all reasonable alternatives
should be explored.  One proposed idea, involving a fundamental
acceleration scale $a_0\,{\simeq}\,1$--$2\times10^{-10}{\rm m}\,{\rm s}^{-2}$,
is called MOdified Newtonian Dynamics or MOND.
MOND was suggested to explain the flat rotation curves of galaxies without
the need for dark matter.  Whether or not it can adequately fit the
available data has been debated for almost 20 years (and we summarise many
of these studies), but only recently have there been studies attempting to
extend MOND to larger
scale regimes.  We discuss how the basic properties of MOND make it at best
ambiguous to apply these ideas to cosmological scales.  We emphasize the
difficulties inherent in developing a full theory in which to embed the
main MOND concepts.  Without such a theory there is no obviously consistent
way to discuss the early Universe and the growth of perturbations.  Recent
claims that MONDian cosmology works very well are therefore not supportable.
We also provide an argument for why $a_0\,{\sim}\,cH_0$ naturally, a
coincidence which is often suggested as a motivation for taking MOND seriously.
We discuss other alternative theories of gravity concluding, as others have,
that no metric theory extensions appear workable for explaining rotation
curves as well as other observed phenomena.  The whole premise of many of
these attempts is fatally flawed -- galaxies are not pre-selected, discrete,
isolated regions which formed monolithically and around which one can
construct an axially-symmetric dynamical model in order to remove the need
for dark matter.  In the modern view, galaxies are part of a dynamic continuum
of objects which collectively make up the evolving large-scale structure of
the universe.
\end{abstract}


\section{Introduction}

A model based on the growth of small fluctuations through gravitational
instability in a universe with cold dark matter (CDM) provides an excellent
fit to a wide range of observations on large scales, ${\ga}\,1\,$Mpc
(see e.g. Liddle \& Lyth~1993, Peacock 1998).
However, the nature and properties of the CDM, apart from its being cold
and dark, remain mysterious.
Since there is no {\it direct\/} detection of this form of matter, 
one should be cautious about accepting the idea of dark matter casually,
and remain open minded to other possibilities.
Recently in fact a great deal of attention has been focused on some apparent
failings of the CDM model on scales of galaxies (e.g. Hogan \& Dalcanton~2000,
Sellwood \& Kosowsky~2000, Firmani et al.~2001).
This examination of details of the CDM model has led to a 
resurgence of interest in other concepts for modelling the dynamics of 
galaxies. 
Ideas which have been proposed to remedy these perceived problems
include self-interacting dark matter, warm dark matter,
fine-tuned initial conditions, and modifications to gravity.
Obviously there are less exotic remedies as well, for example those
depending on baryonic processes, such as gas physics, cooling and feedback
mechanisms.  Baryonic contributions to dynamics
are in many cases expected to be non-negligible but are as of yet
notoriously difficult to calculate reliably.

The focus here will be specific proposals to modify gravity, which
has been the subject of much attention of late.  It has been suggested
that the model for the fundamental laws of physics could be changed to 
accommodate astrophysical observations of galaxy dynamics without the need 
to invoke dark matter.
The best known such example is Milgrom's theory of Modified Newtonian
Dynamics (MOND; Milgrom~1983a, 1994).  The body of literature on MOND
ranges from bold claims of success in every tested arena 
\cite{Milgrom99} to 
discussions and identification of many of its failures
(e.g.~van den Bosch \& Dalcanton~2000a).
While it is clearly prudent to investigate all the alternatives to the dark
matter hypothesis, particularly since no specific candidate particle has been
detected (except perhaps a close to negligible neutrino contribution),
we conclude in this paper that MOND does not appear to be promising.

Much of the focus of MOND has been on the study of galaxy dynamics and
structure.  We shall briefly review this work and its criticisms below.
Recently, however, there has been a push to extend MOND to the cosmological
realm (e.g.~McGaugh~1999, 2000, Sanders 1998, 2000),
and it is this aspect of MOND which is our main concern.  One claim is that
the low amplitude of the second acoustic peak in the CMB anisotropy data
is a `prediction' of MOND.  In fact as we will discuss, this calculation is
simply for a high baryon density model dominated by a cosmological constant,
which also fails for other reasons.  But in fact there is no obviously
consistent way to carry out such calculations in a truly MONDian picture.
We find several difficulties with MONDian cosmology that appear to be
insurmountable.  Avoiding all of these difficulties makes the MOND picture
appear more and more `epicyclic'.  The existence of some form of non-luminous
matter within the context of conventional physics seems, to us,
considerably simpler.

We start by listing many of the motivations for dark matter in the
standard model,  We then enter into a discussion of the failings of
baryons-only predictions
for one of the main pieces of cosmological data, namely Cosmic Microwave
Background (CMB) anisotropies.  This is relevant, since the predictions for
MOND and standard cosmology {\it might\/} coincide for this
calculation \cite{McGaugh00}.
We briefly describe MOND and then delineate
its conceptual and empirical difficulties, including a review of 
the numerous issues previously discussed by other authors.
We continue with a discussion on structure formation in these models.
We then mention other alternatives which share some of the features of
MOND, as well as some additional
future tests which will soon be possible.

\section{Dark matter in the standard model}

\subsection{Evidence for Dark Matter}
Let us begin by reviewing the main evidence for large amounts of matter
in the Universe which are not associated with the luminous components.
There are several observations which are usually
interpreted as providing evidence for (cold) dark matter, including:
\begin{itemize}
\item the rotation curves of galaxies compared with their light distributions;
\item the gas content of clusters compared with velocity, x-ray or lensing
mass estimates;
\item the normalization of galaxy clustering compared with microwave
anisotropies;
\item the shape of the large-scale galaxy correlations;
\item the lack of strong Silk damping and existence of small-scale structure
(e.g.~Ly-$\alpha$ forest);
\item cosmic flows and redshift space distortions;
\item and the amplitude of weak lensing by large scale structure.
\end{itemize}
Recent reviews of several of these topics are provided by e.g.~Dekel,
Burstein \& White~(1997), Bosma (1998),
Turner (1999), Peacock (2000) and Primack (2000).

This list is not exhaustive.  Let us give another, more recent example,
coming from
CMB experiments probing the damping tail (the Cosmic Background Imager
experiment in particular)
which provide a further independent constraint on $\Omega_{\rm CDM}$ and
$\Omega_{\rm B}$
(see White~2001 for more extensive discussion). 
Here $\Omega_{\rm CDM}$ and $\Omega_{\rm B}$ are the density parameters in
cold dark matter and baryons, respectively, and $\Omega_{\rm M}$ is the sum
of the two.
The ratio of the damping scale and the spacing between the acoustic peaks,
$\ell_{\rm D}/\ell_{\rm A}$, depends only on $\Omega_{\rm M}h^2$ and
$\Omega_{\rm B}h^2$.
It is independent of the distance to last scattering, i.e.~the geometry of
the Universe and any late time effects such as a cosmological constant or
quintessence.  The sense of the $\Omega$ dependence is
that, at fixed $\Omega_{\rm B}h^2$, the lower the
matter density the smaller is $\ell_{\rm D}/\ell_{\rm A}$.
Phrased another way, if we require a peak at $\ell\,{\sim}\,200$
then the amount of
power at $\ell\,{\sim}\,10^3$ decreases exponentially as we lower
$\Omega_{\rm M}h^2$.
Thus lower limits on $C_{\ell}$ for $\ell\sim 10^3$ can provide strong
lower limits on the physical matter density of material in the Universe
in an almost model independent way.

\begin{figure}
\begin{center}
\leavevmode
\epsfxsize=8cm \epsfbox{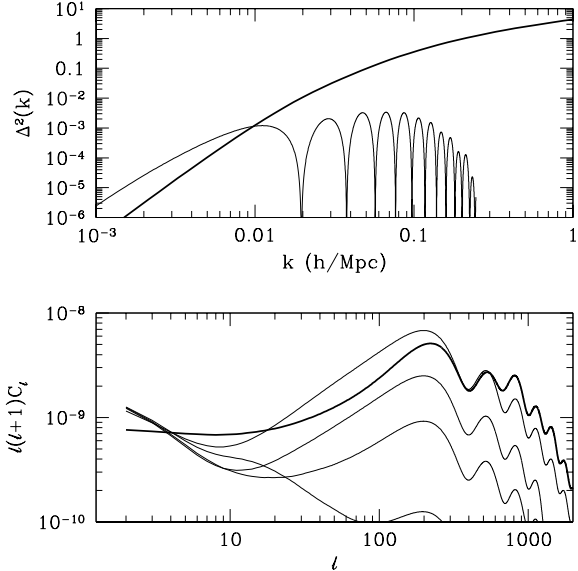}
\end{center}
\caption{Power spectra for (matter) density perturbations (upper panel)
and for CMB anisotropies (lower panel).
The thicker lines are the usual $\Lambda$CDM `concordance'
model, while the other lines are the vacuum-dominated ($\Lambda$BDM) model
suggested by McGaugh~(2000).  All models have been {\sl COBE\/}
normalized.  In the lower panel the four $\Lambda$
curves are for reionization optical depths of $\tau=0$, 0.5, 1 and 2, with the
latter perhaps the most likely for a MOND picture in which galaxies form
early.  The MONDian optimistic view would be that all of the shortcomings
of these curves could somehow be solved in a full cosmological theory which
includes modified gravity.
}
\label{fig:power}
\end{figure}

MOND represents an attempt to explain the first item in the above list
without the need for dark matter, and there are claims that some of the
other items can also be accommodated (see, e.g., Milgrom~1999).  
There is clearly a difference of opinion
here among researchers about how to apply Occam's razor to the dark matter
question.  Some authors appear to regard as anathema the idea that some matter
might not be very luminous.  We suggest that it might indeed be reasonable
to consider radical changes to well-known physics {\it if\/} rotation curves
were the sole piece of evidence in favour of dark matter.  But this is far
from being the case.  A baryons-only universe actually has difficulties on
several fronts, as we now discuss.

\subsection{A baryonic universe?}
The models discussed by McGaugh (2000) and Sanders (2000) have relatively
high $\Omega_{\rm B}$, and only baryonic dark matter, apart from a large
contribution from $\Omega_{\Lambda}$, to make the Universe flat or somewhat
closed.  These authors then appeal to possible physics within a MONDian
cosmology to fix up some of the immediate problems.

Before discussing the MOND proposition itself, we will mention two
consequences of having no CDM in the usual paradigm.  These two
particular consequences
depend only on the cosmological parameters used in the MOND cosmology
and not on MONDian dynamics per se; as they are more general, we mention them
before tackling MOND more explicitly below.

In order to fit recent CMB data (see e.g.~Pierpaoli, Scott \& White~2000,
de Bernardis et al.~2000, Hanany et al.~2000),
a model with no CDM has to
have a baryon fraction which is at least mildly challenging for
nucleosynthesis constraints, and it requires a cosmological constant which 
is {\it much\/} larger than lensing and other constraints allow.  
The number of strong gravitational lenses for $\Omega_\Lambda\,{\sim}\,1$
would be over 5 times the currently observed number \cite{Kochanek}.
Since the effective value of $\Omega$ contributed by stars and gas in known
components of galaxies is estimated to be around 0.004 (Fukugita, Hogan \&
Peebles~1998), it is also true that McGaugh's model needs to contain about
an order of magnitude more dark baryons than luminous baryons.  Therefore
even if particle dark matter is avoided, this is not true for dark matter
in general.

It should be noted that the CMB fit itself, for the
simplest case, is also poor.
Fig.~1 shows that {\sl COBE}-normalized models with only baryons (and a
large cosmological constant component in order to make the Universe close
to flat) has slightly too high a peak -- these models were discussed in
McGaugh (2000) but this difficulty was not stressed.
Hence in order to use baryons alone to fit CMB data like those from the 
BOOMERANG experiment, it is
necessary to invoke some other physical explanation for why the large
angle anisotropy signal has been underestimated relative to the smaller scales.
This may be possible to
achieve by changing the model from the simplest case, for instance
using a combination of gravity waves, tilt and reionization.  Naively
the modification required to make the spectrum fit are quite severe and
arguably fine-tuned.  In order to quantify the difficulties in utilizing
reionization, more details of structure formation in the baryons-only theory 
are needed.  Moreover, the matter power spectrum (upper panel of Fig.~1)
is disastrous for baryon-only models.

Combining the CMB and any one of a number of other cosmological constraints 
makes the high $\Omega_{\rm B}$ solution untenable.  This was stressed by
Lange et al.~(2000), Jaffe et al.~(2000) and Bond et al.~(2000) in direct
analysis of BOOMERANG and Maxima data, as well as through careful
multi-parameter fits by Tegmark \& Zaldarriaga (2000) and others.
There are two disjoint regions of parameter space that can with more or
less equal capacity fit the CMB data {\it taken alone}.  But the high
$\Omega_{\rm B}$ region rapidly shrinks to zero when other constraints
are added.
Griffiths, Melchiorri \& Silk (2001) use the CMB data plus supernovae
observations to conclude that no reasonable baryon-only model can fit.  In
fact the no-CDM part of parameter space is restricted to closed models with
$\Omega_{\rm B}\simgt0.13$, and this is discrepant with Big Bang
nucleosynthesis limits (as well
as implying an amount of dark baryons almost as high as the amount of
CDM required normally). 

The comparison of the CMB amplitudes with a galaxy clustering normalization,
such as $\sigma_8$ ( the variance on $8h^{-1}$Mpc scales)
also pose problems for this model.  As these
arguments relating the CMB to something like $\sigma_8$ rely on a
number of assumptions, it is prudent to check for possible loopholes.
Three possibilities certainly come to mind: (1) $n>1$ power spectra, or
power spectra with features in fortuitous places; (2) the existence of an
extra dark matter component, such as hot dark matter; or (3) extra fluctuations
due to an isocurvature component or topological defects.  Such modifications
are not going to simultaneously solve all the problems inherent in the
power spectra shown in Fig.~1, however.

Since the $\Lambda$CDM model shown in Fig.~1 is a fairly good fit to all
the available data, both for the matter fluctuations $\Delta^2(k)$
($\equiv 2\pi k^3 P(k)$) and the CMB anisotropies $C_\ell$, it can be seen that
high $\Omega_{\rm B}$ cosmologies within the standard framework are
an extremely poor match to the observations.  
If CDM is to be avoided at
all costs, then the only way to go is to be much more radical about the
cosmological framework.  The hope might be that, with sufficiently drastic
changes, some of the shortcomings of high $\Omega_{\rm B}$ models might
be avoided.
We now turn to MOND, one such radical modification of gravity.

\section{MOND}

The best known suggested 
modification to Newtonian gravity is usually referred to
as Modified Newtonian Dynamics or MOND (Milgrom~1983a,1983b,1983c).  
Similar ideas for
avoiding the need for dark matter been put forth by several other authors
(e.g.~Kuhn \& Kruglyak~1987, Bekenstein~1988, Mannheim \& Kazanas~1989,
Liboff~1992).
It is difficult to be precise about the MOND idea, because the literature
does not form a coherent whole.
Fundamental to the idea of MOND is that it is an `effective' theory,
playing a role similar to Kepler's laws (as stressed by Felten~1984).
The proponents of MOND have yet
to develop the analogue of Newtonian mechanics to explain this effective
theory.  The absence of a full theory seriously limits the predictive power of
MOND, and leads various authors to disagree as to what the observational
consequences of this revision will be.

The basic concept behind MOND is that there exists a fundamental acceleration
$a_0$ ($\simeq2\times 10^{-10}{\rm m}\,{\rm s}^{-2}$ with an uncertainty of
a factor $\sim 2$), {\it below\/} which the actual acceleration is
{\it larger than\/} the Newtonian one.
This is sometimes formulated by relating the `Newtonian' acceleration,
$a_{\rm N}$, of a test particle to the actual observed acceleration, $a$,
through the relation (Milgrom 1983; Bekenstein \& Milgrom 1984)
\begin{equation}
\label{basicmond}
  \vec{a}_{\rm N} = \mu(a/a_0) \vec{a},
\end{equation}
where $\mu(x)$ is an interpolating function with limits
\begin{equation}
  \mu(x) = \left\{
  \begin{array}{ll} x & x\ll 1 \\
                    1 & x\gg 1.
  \end{array}
  \right.
\label{eqn:mudef}
\end{equation}
We shall refer to the limit $a\ll a_0$ as the `MOND regime' and in this
limit
\begin{equation}
  a = \sqrt{ a_0 {F\over m} },
\end{equation}
which is often given as the fundamental equation of MOND.
Milgrom (1983) suggests $\mu(x) = x\left( 1+x^2\right)^{-1/2}$, in which case
a test particle in a spherical potential experiences an acceleration
\begin{equation}
  a = {a_{\rm N}\over\sqrt{2}}
      \left[ 1+\sqrt{1+4\left({a_0\over a_{\rm N}}\right)^2}\ \right]^{1/2}
  \ .
\end{equation}

Alternative suggestions have been made for modifying gravity,
whereby a logarithmic potential is
added to the usual Newtonian one.  Such a formulation can also be cast
in the form of Eq.~(\ref{eqn:mudef}) with
\begin{equation}
  \mu(x) = { \sqrt{1+4x}-1\over\sqrt{1+4x}+1 }
\end{equation}
(Kinney \& Brisudova~2000).  However the additive formulation is not exactly
the same as `classical' MOND, in that it specifies a fundamental
{\it length\/} scale rather than a fundamental {\it acceleration\/} scale.
Furthermore, the additional potential can be made proportional to the
product of the masses of the interacting bodies, thus circumventing some of
the problems we detail below.
We now turn to conceptual difficulties with MOND and
then with MONDian cosmology.  We then summarize empirical difficulties
over a range of scales.

\subsection{Conceptual difficulties with MOND}

While it seems straightforward to modify the universal law of gravitation
on large scales, this is in fact not so.  Newton's laws of motion and the
law of gravitation are closely woven together in such a way that simple
modifications rapidly lead to unpalatable consequences. 

One immediate question is whether MOND applies equally to {\it de\/}celerations
as to accelerations, or indeed whether the motion needs to be just a change
in the vector direction of acceleration in order to show MOND effects.
The usual interpretation is that all changes in velocity are subject to MOND.
Immediately, we see a fundamental difference with standard dynamics when we
consider a test particle moving away from a central mass.  In the MONDian
picture, the test particle's deceleration never drops below the value $a_0$.
Hence it cannot escape to infinity -- in MOND there are no unbound orbits.
We will return later to this conceptual property of MOND, which is joined
by several other ambiguities which make it difficult to outline a consistent
MOND picture when multiple systems are involved.

Several profound difficulties with Milgrom's original proposal as stated
were identified by Felten \shortcite{Felten84} 
soon after the introduction of the MOND prescription. 
One example is that MOND violates Newton's third law, since it is
not symmetric between a galaxy and a test mass.  
Another way of stating this is that because acceleration is
not inversely proportional to mass,
momentum is not in general conserved for an isolated system.  In addition,
since the gravitational force is no longer linear, the simplest MOND
incarnations have the total gravitational force not being equal to the sum
of the partial forces.  In particular, the motion of the centre of mass of a
body no longer obeys\footnote{As Felten~(1984) points out, this makes the
original papers of Milgrom inconsistent, or ambiguous at best,
since astronomical data describe multi-particle systems.}
the familiar motions of undergraduate mechanics unless
all of the bodies in the system have equal masses.

One can still go forward with numerical studies by implementing some
prescription for determining accelerations in a given system.  However,
on a technical level, this property makes simulating a MOND system extremely
challenging, as techniques such as N-body cannot be applied unless each body
represents an actual physical component of the system under study
(e.g.~an individual star or perhaps even an atom).
As dynamical friction within the context of
MOND will be different, basic interactions such as merging will also
undergo modification.

By specifying a fundamental constant with the dimensions of acceleration,
MOND implicitly violates Lorentz invariance, and thus cannot be fit into our
conventional theories of modern physics.
Coordinate invariance, which underlies relativity,
is gone, as is any concept of modelling forces as exchange particles.

Much of the challenge involved in making MOND part of a complete and
consistent theory is due to the realm where it extends Newtonian theory.
One often thinks about how a more complete theory gives a {\it stronger\/} 
effect than the simpler theory in the strong regime -- but for MOND the 
acceleration is {\it larger\/} (than in the Newtonian theory) when it 
falls {\it below\/} a particular value.  MOND must have this sign in order to
explain rotation curves with no dark matter.  However, this is the source
of many of the problems at large scales.

A more specific problem is that the MONDian `potential' tends to $\log(r)$ at
large distances, as the Newtonian acceleration, given a large enough distance,
will always fall below the MONDian critical value.
So unless the potential is further adapted to cut it off somehow, then
all objects are bound to each other; there is no such thing as escape
velocity in MOND.  There are cosmological extensions of MOND which
address this and these will be discussed below.

The biggest theoretical dilemma comes in deciding how to
interpret accelerations when multiple motions are involved: do we consider
some absolute acceleration, or only peculiar accelerations?  We will
elaborate on this important point below.  Suffice to say that this makes it
hard to place MOND within a cosmological context, and suggests that the
Cosmological Principle cannot really apply.

This difficulty deciding {\it which\/} acceleration one considers was already
understood by Milgrom (1983a) when he proposed MOND.  This arises from
the problem in explaining the lack of a mass discrepancy in wide
binaries \cite{Cloetal} and open clusters \cite{LeoMer}.
Considering orbits around the Sun, MOND effects would become important for the
solar system at about $7000\,$AU, which is much closer than the nearest stars.
Hence, in loose groups of stars, one might expect to find that stellar
evolutionary models for the stars would require less mass than implied
dynamically by the groups.  This has been avoided in Milgrom's prescription
by choosing a preferred frame and considering the {\it absolute\/}
acceleration in this frame.  However, this would mean throwing out some
fundamental
physics concepts which underpin ones ideas about relative motion.
The proposal is that if a group of stars is accelerating
around the centre of the Galaxy, and that acceleration is larger than $a_0$,
then there will be no MOND effects observed, even if the relative accelerations
among the stars are less than $a_0$.  In other words MOND
essentially requires an absolute meaning for acceleration, and a special
status for galaxy-scale collections of mass.
We will come back to this when we try to understand MOND within the
cosmological context in the next sub-section.

In summary, here are some of the concepts naturally implied by MOND, which
would have to be tackled to make it a complete theory:
\begin{itemize}
\item MOND explicitly violates the equivalence principle;
\item MOND violates conservation of momentum (Newton's third law);
\item MOND violates Lorentz invariance;
\item MOND may violate the Cosmological Principle;
\item MOND does not allow for superpositions of gravitational fields;
\item and MOND suggests that all bodies are bound to each other.
\end{itemize}
Thus we see that MOND is far from being simply a modification of a law
in a poorly tested regime.  It violates basic principles which underlie
our entire framework for theoretical physics.
This is a very high price to pay for explaining even the most
puzzling of astronomical data!
It may be that some of the obstacles listed above can be overcome if MOND
could be embedded within some complete and consistent theory.  However, one
should not underestimate the magnitude of this task.

Although there have been attempts to derive full theories which have
similarities with MOND, these have not been very successful.
Bekenstein \& Milgrom (1984) presented a toy model for a full theory of
gravity which might contain MOND.
This model gets around some of the problems described above by defining the
force to satisfy $F=ma$, but at the end of the day is simply a non-relativistic
potential model that falls far short of predicting the wider behaviour of such
a theory.
Mannheim \& Kazanas (1989, see also Mannheim 1997, 2000, and references
therein) attempted to derive a covariant general theory of gravity that
has some features in common with MOND.
Sanders (1997) described scalar-tensor theories that might accommodate MOND.
However, MOND certainly implies a preferred frame and leaves details
of how to calculate anything still very unclear.  Typically such ideas
applied to the scale of galaxies (for example) treat the centre of the
galaxy as a special point about which to perform calculations.
Similar bi-metric theories
have also been proposed by other authors (e.g.~Drummond 2001 and references
therein).  None of these models
appears `natural', all suffer from further physical awkwardness (e.g.~causality
problems, behaviour of local gravity, stability considerations
or gravitational lensing), and none provides a detailed
framework in which to carry out cosmological calculations.

Periwal (1999) suggests an approach to quantum gravity involving an ultraviolet
(i.e.~high energy)
fixed point which might have non-Newtonian dynamical effects at large scales.
The basic idea is
that if we assume the existence of an ultraviolet fixed point for gravity 
then there will exist a scale at which gravity will become a strong force.  
This mechanism operates in quantum chromodynamics
where the force is `weak' at short distances but becomes
strong enough to confine quarks into hadrons at long distances.  In analogy,
gravity might exhibit qualitatively different (and perhaps stronger) 
behaviour at very long distances with a characteristic length scale $\xi$ or 
acceleration scale $\sim 1/\xi$.  
To make this proposal more concrete, one would like to
to compute the existence of the fixed point, the scale at which the transition
takes place and the expected qualitative changes, however there is
not yet currently a calculable theory of quantum gravity.

Others have suggested ideas with similar flavour to MOND, but typically
these are phenomenological models only, for example introducing a special
scale rather than an acceleration, or modifying $G$.  None of the suggestions
that we are aware of help with the issue of how the concept behind MOND
might be embedded in a genuine theory.

\subsection{Difficulties with MONDian cosmology}

To say that we have no idea how the early Universe works in a MONDian picture
would be a gross understatement.
The fact that MOND is not relativistic makes it difficult to interpret
the scale factor and the Hubble law, and the possible acceleration of the
Universe.  In order to perform cosmological calculations, one needs
some way of generalizing the theory.  

To describe cosmology, we will assume that we start with homogeneous and 
isotropic initial conditions.  
It is not obvious how to produce scale invariant initial conditions such
as would arise from inflation in such a theory, however we will focus on
consequences from MOND for subsequent structure formation here.
Specifically, we imagine that somehow a theory
can be found which at early times
results in a universe that is, to a good approximation, homogeneous and
isotropic.  Since we must be careful not to make relativistic statements in
this theory, we shall assume that this holds true in the `absolute' rest
frame of the Universe (we shall return to this below).

We then need to include the Hubble expansion and perturbations. 
The most explicit work on cosmology within MOND is by Sanders (1998, 2000),
using ideas explored earlier by Felten (1984), and we will describe
their generalizations below.  The possible cosmological prescriptions
differ in their applications of the original MOND equation,
$\mu(a/a_0)a=a_{\rm N}$, distinguished by
which accelerations are used in which place on the left hand side.
Separation is made into a `peculiar'
acceleration $a_{\rm P}$ and a `Hubble' acceleration $a_{\rm H}$.
The Hubble acceleration is taken from the Friedman equation
\begin{equation}
\frac{\ddot{R}}{R} = - 4 \pi G (\rho + 3 p),
\end{equation}
with a point at position $r = R(t) r_0$ having acceleration
$\ddot{r} = r {\ddot{R}(t)}/R(t) r$.  Note that the acceleration
thus depends on the reference point used to determine $r$ (that is,
$r_0$).

Requiring the MOND prescription to affect $a_{\rm P}$ at the
very least, in order to reproduce effects in galaxies, we have
the following four possibilities:
\begin{enumerate}
\item $\mu( (a_{\rm H}+a_{\rm P})/a_0 )(a_{\rm P}+a_{\rm H}) = a_{\rm N}$;
\item $\mu(      a_{\rm P} /a_0 ) a_{\rm P}                  = a_{\rm N}$;
\item $\mu(      a_{\rm H}/a_0 ) a_{\rm P}                   = a_{\rm N}$;
\item $\mu( (a_{\rm H}+a_{\rm P})/a_0 ) a_{\rm P}            = a_{\rm N}$.
\end{enumerate}
One could also imagine $a_0$ varying with cosmological time, a suggestion
which has already been made (e.g.~Sanders 1998),
and which we do not consider further (but see \S4).  We reiterate that
that {\it both} accelerations $a_{\rm P}$ and $a_{\rm H}$ 
implicitly assume a
specific frame of reference and thus further specification is needed to
complete these definitions.

Now let us consider each of these four possibilities in turn.

It is simplest to consider the first option (i) for the case of pure Hubble
expansion (i.e.~set $a_{\rm P}=0$, which will be true at sufficiently
early times).  Choosing the `absolute' rest frame as described above, with
this prescription, initially
the total acceleration on any fluid element tends to zero
(the MOND regime) because of the homogeneity and isotropy.  The vector sums
of all of the accelerations tend to zero simply using spherical symmetry about
any point.  Note that this argument is completely independent of the functional
form of the force law, it depends only upon the symmetries of the problem and
that the force acts along the line joining the two bodies (so as to have for
example Kepler's equal area law for elliptical orbits).

This is the case considered by Felten (1984),
in analogy with the McCrea \& Milne~(1934) approach to deriving the Friedmann
equations from Newtonian dynamics.
As noticed immediately by Felten (1984), the `cosmological equations' of
MOND do not admit a homogeneous and isotropic universe that obeys the
cosmological principle.
As distant objects in MOND remain bound to each other with an
approximately logarithmic potential, perturbations in the Universe will
presumably\footnote{Since it is at present impossible to formulate a
self-consistent cosmology, it is not possible to prove that perturbations
will always collapse.} always collapse -- every system will have negative
total energy.  Assuming a central point can be chosen somehow, regions
collapse quickly, out to scales of perhaps $30\,$Mpc in
the present universe (Sanders~2000).  
For this reason, Sanders makes the assumption that MOND only
applies to determining the peculiar accelerations.  This
allows the next three prescriptions.

We note that another way
to violate this argument is to suggest that the frame in which the Universe
is isotropic and homogeneous is in fact accelerating with respect to some
fundamental frame against which we measure accelerations in the MOND universe.
If this is true then there is {\it no\/} scale on which MOND effects operate.
We could turn MOND back on by having the entire Universe decelerate with
respect to the fundamental frame at a later time, in effect `turning on' the
MOND force.  Obviously this mechanism can allow us to turn MOND on and off
as many times and at whatever points we desire.
Note however that in the transition regime, where the acceleration of the
Universe with respect to
our background frame is ${\cal O}(a_0)$, there will be a preferred
direction to the force law.  Objects whose matter induced acceleration is
oppositely directed to our fundamental acceleration will have more `MOND
acceleration' than those which are accelerating in the same direction.  The
existence of a preferred direction would be quite an unpleasant side-effect of
this mechanism.

Let us now turn to case (ii) from the above list,
As soon as perturbations are introduced into our early Universe,
MOND will kick in with full force and the evolution will be distinctly
different than in the conventional theory.  Hence CMB anisotropy
calculations using {\sc cmbfast} (Seljak \& Zaldarriaga 1996)
for example, are not valid.  And, in general, there is no way to get anything
like the usual cosmological results in this case.

For (iii), the opposite problem occurs. Again, a 
central point for the Hubble flow must be chosen
and any system far enough away from it will again show no effects of
MOND.  Thus MOND would only apply within a certain distance of some
specified region.  

Option (iv) is the one advocated by Sanders (2000).
To go further, one has to decide how to define the peculiar
and Hubble accelerations.  The first option is to choose one
point and to define these accelerations from this position.
This quickly has the same problem as suggestion (iii).  That is,
we take as a physical model, for example, that the Universe is
empty and large, and into this universe explodes an expanding fireball
containing all of the matter and radiation in what we think of as our Universe.
The fireball cools and  the expansion (rapidly) slows due to self-gravity.
It is this overall acceleration that stops us from feeling the MOND force.
We fortunately live at the centre of this expanding fireball which has now
cooled to the point where we are matter-, and not radiation-dominated.
Thus the McCrea analogy for deriving the Friedmann equations must in fact
represent the true physical situation and not be a convenient pedagogical
way of invoking Birkhoff's theorem.

However, if one proposes an absolute meaning for acceleration,
then a distant galaxy or cluster has a large $a_{\rm H}$ and so its internal
kinematics would be Newtonian not MONDian, and hence this would not explain
rotation curves, etc.  The key point is that you cannot imagine
moving to the rest frame of that distant galaxy (which would allow it
to have a flat rotation curves) because you have given an absolute meaning to
acceleration in MOND.

A second way to address this appears to be the intention of
Sanders (2000) -- one considers a shell which
has broken away from the general expansion.  In this case, $a_{\rm H}$
and $a_{\rm P}$ are defined with respect to the centre of this collapsing
shell.  This means assuming an isolated overdensity, 
insensitive to other possible sources of peculiar velocities,
i.e.~other inhomogeneities outside of it.
This seems hard to reconcile with the modern cosmological view.
In the conventional picture, velocities are
driven by perturbations over a range of scales, and tidal forces generate
galaxy spins, for example.
Nevertheless, if we assume this is somehow true, we still run into
difficulties.  At the same time as ignoring forces from external
perturbations, the background cosmology can 
affect the expansion of this shell -- in fact it decides the transition
between the Newtonian and the MONDian regime.

It becomes awkward to try to apply this rule to each
shell in the specified region, as accelerations cannot be separated in MOND and
thus another ambiguity arises.
One possibility is that one applies this rule to the largest such 
shell in each region.  Thus, the Hubble acceleration for scales in the shell 
is taken to be whatever the Hubble
acceleration is at that scale measured from the centre of the largest
collapsed shell containing the system.
At the present time, the largest such shell would be approximately
cluster-sized and so perhaps MOND should not be used for any galaxies on
the outskirts of a cluster?

A problem with this definition is that a region will go from 
the Hubble flow to the MOND region and then perhaps out of the MOND
region all depending on its distance from some central point.  The
definition of this central point will often change in the process of merging
and collapse of larger scales in what appears to be an instantaneous
non-local manner.  Nevertheless,
this prescription seems closest to what is suggested in Sanders (2000).
However the ambiguities involved make it difficult to make any definitive
statements about how to treat cosmological perturbations within MOND.

In addition, the problem above, that the Hubble acceleration can be $\leq a_0$,
will also have been expected to occur in our history as the Universe
is now accelerating but was at one time (in both the concordance cosmology
and e.g. that suggested by McGaugh~2000) decelerating (we return to this in
\S3.4).

If we proceed with the calculation using the Hubble flow as part of the
acceleration for purposes of invoking MOND (option (iv) above), then there
is an ambiguity about {\it which\/} scale we are supposed to use.
In other words, the outer regions of some galaxy could define the relevant
shell and then we would say that MOND should be used for that particular
galaxy.  But we could also look at a nearby galaxy and say that it is
part of a bigger shell, in which case we do not use MOND. 
How one would consistently treat a set of neighbouring merging shells,
destined to become a filament for example, is entirely unclear.

Sanders (2000) deals with the case of one collapsing shell, and he
agrees this prescription as it stands is a first approximation and
possibly not self-consistent.  The growth of non-linear structure, 
which involves collapsing objects within other collapsing objects and 
their collisions, is thus difficult to calculate reliably.  These inherent
ambiguities, together with the smallness of the Hubble acceleration, and the
intrinsic non-locality of this picture, make option (iv) seem quite
unworkable.

So where have we come in this discussion?  If it has appeared less than
crystal clear, then it is because MOND is both ambiguous and not obviously
self-consistent when it comes to extending the idea beyond individual galaxies
to the cosmological context.  There are several possibilities for {\it which\/}
acceleration one should applies the MOND equation.  We have tried to consider
each possibility in turn.  The conclusion is that it is difficult to find
any consistent framework in which calculations could be attempted.  It seems
that the only way to have MOND make sense is to arrange for the centres of
galaxies to be pre-selected as special places around which to consider
MONDian gravity.  Perhaps this would be reasonable in a picture where galaxies
were discrete, isolated and formed monolithically.  However, it is very hard
to reconcile this with the Universe as presently understood, containing
fluctuations over a wide range of scales, as well as merging hierarchical
clustering and the rest.

\subsection{Empirical difficulties with MOND}

Some authors have indicated that MOND has been very successful in
explaining observations of rotation curves for a variety of objects over
a wide range of scales (see e.g., Milgrom 1999).  
We wish to stress that a number of studies have
indicated difficulties in reconciling MOND with data under the assumption
that there is {\it no\/} dark matter, and that the scale $a_0$ has a fixed
value.
Dressler \& Lecar (1983, cited in Felten~1984) appear to have been the first
to object that MOND did not adequately fit their data, but they did not
publish this.  Subsequently, a large body of published work has
accumulated over the years.  Since this corpus is not
referred to very much in the MONDian literature, let us try to be fairly
comprehensive here.

Kent (1987) pointed out that although MOND could fit his H\,{\sc i}
rotation curve
data there was a factor of 5 range in the value of $a_0$ required and also
no clear evidence for the slightly falling rotation curves that MOND would
still predict.
Hernquist \& Quinn (1987) examined simulations of shell galaxies within
MOND, concluding that the observed number and radial distribution of shells
in NGC 3923 could not be explained without a dark matter halo.
The \& White (1988) found that a MOND fit to the Coma cluster requires
a higher value of $a_0$ than for galaxies and also does not predict the
correct temperature profile for the x-ray emitting gas.
Lake (1989) pointed out discrepancies between MOND and observations of
dwarf irregular galaxies.
In particular Lake \& Skillman (1989) found that MONDian fits to the Local
Group dwarf IC~1613 would require values of $a_0$
at least an order of magnitude below the favoured values.
Kuijken \& Gilmore~(1989) also considered MOND in their study of the
distribution and dynamics of K dwarfs.
They concluded that the vertical accelerations in the solar neighbourhood
require the presence of a dark halo and are quite severely inconsistent with
MOND.  Gerhard \& Spergel (1992) investigated dwarf spheroidal galaxies in the
Local Group and found that some of the dwarfs need to contain some dark
matter even under the MOND hypothesis.
Gerbal et al.~(1992, 1993) compared data on several x-ray clusters with MOND,
finding that in several cases the MOND fits suggested less mass than implied
by the x-ray emitting gas alone.  In addition MOND predicted too strong a
concentration of the mass towards the centre, and could not in general fit
the data without requiring dark matter.
Lo, Sargent \& Young~(1993) described neutral hydrogen data for dwarf irregular
galaxies which would have a mass below the observed H\,{\sc i}
mass under the MOND
hypothesis.  Christodoulou, Tohline \& Steiman-Cameron (1993) studied tilted
ring models for the H\,{\sc i}
distribution in NGC5033 and NGC5055.  They found
evidence for prolate potential wells, which would be difficult to accommodate
in MOND.
Buote \& Canizares (1994) discussed the ellipticities of the x-ray vs optical
isophotes of the elliptical galaxy NGC 720 showing that MOND cannot explain
the flattening of the x-ray isophotes without requiring a component of dark
matter.
Soares (1996) found that MOND can only explain the velocity
distributions of binary galaxies if either the mass to light ratios are high
(i.e.~there is lots of dark matter) or the galaxies are on very eccentric
orbits, and in that case the distribution of separations is not consistent.
S{\' a}nchez-Salcedo \& Hidalgo-G{\' a}mez (1999) looked at disc
instabilities of gas-rich dwarfs within MOND and concluded that these would
be catastrophic.
Giraud (2000) discussed how variations in mass-to-light ratios between galaxies
in the standard dark matter picture
are more palatable than variations of $a_0$ within MOND.
Blais-Ouellette, Amram \& Carignan~(2000) reported
problems fitting the detailed kinematics of
the well-studied late type galaxy NGC 3109 within MOND.  Additionally they
found that fits to late type galaxies
appear to require higher values of $a_0$ than for early type galaxies.
Bothun et al.~(2000) investigated a new sample of Cepheids in the large spiral
NGC 2841.  MOND could give an acceptable fit to the rotation curve of this
galaxy, but only if it lies considerably further away than its Cepheid-implied
distance.  van den Bosch \& Dalcanton~(2000a, 2000b) 
have carried out a very detailed study of
dwarf rotation curves in the MOND picture.  They concluded that MOND cannot
reproduce both the Tully Fisher relation and the lack of
high surface brightness dwarf galaxies.
Dalcanton \& Bernstein~(2000) have continued the detailed study of galaxy
rotation curve data.  They found that MOND fits to low surface brightness
galaxies require mass-to-light ratios which are too small to be consistent
with stellar population models.

Detailed dynamical investigations of galaxies which argue against maximal discs
(e.g.~Courteau \& Rix~1999) are by inference also arguing against MOND.
In addition, studies which require haloes to be much less flattened than the
luminous distribution will also tend to be inconsistent with MOND. 
Two recent examples of such studies are the persistence of tidal streams in
the Milky Way halo (Ibata et al.~(2001) and the statistics of two- and
four-image gravitational lenses (Rusin \& Tegmark~2001; though other factors
could also be at work in this analysis).

\subsection{MOND at large scales}

The most recently discussed MONDian cosmological model (McGaugh~2000)
is a Friedmann-like universe with a large cosmological constant
($\Omega_{\rm B}\,{\sim}\,0.04,
\Omega_\Lambda\,{\sim}\,1$).
We have already discussed (in \S 2.2) some of the consequences for those
scales where MOND is not operative and where one can apply constraints on
the MONDian choices of cosmological parameters.
We now turn to the MONDian cosmological consequences, beginning with a picture
closest to the conventional cosmology, and then introducing the
extensions described by Sanders (2000), assuming that somehow a consistent
framework could be developed.  Recall that, in this case, the expansion of the
Universe contributes an `absolute' acceleration that must be considered
when evaluating MONDian forces.

In the standard cosmology, the Friedmann equations in a flat Universe give
the acceleration of the scale factor as
\begin{equation}
  {\ddot R\over R} = H_0^2
  \left(\Omega_\Lambda - {1\over 2}\Omega_{\rm M}\right).
\end{equation}
We shall follow McGaugh~(2000) and assume this is the acceleration in a
MONDian universe also.

Let us first consider an object at rest in comoving coordinates, neglecting
look-back time effects.  We then have for an object at distance $d$,
\begin{equation}
  a = 3.24\times 10^{-13} \left( {d\over 1\,{\rm Mpc}} \right)
      h^2 \left( \Omega_\Lambda-{1\over 2}\Omega_{\rm M}\right)
      {\rm m}\,{\rm s}^{-2}.
\end{equation}
If we use interpretation (iv) and choose one reference point,
difficulties arise immediately.  We list these and then go to
the modification (described above) implemented by Sanders (2000).
With one reference point for accelerations, 
for the standard $\Lambda$CDM model, we find that the MOND acceleration is
reached at a distance of approximately $600\,h^{-2}$Mpc from us.
In the close to flat, cold dark matter-less $\Lambda$BDM models proposed by
McGaugh (2000), this value is more like $400\,h^{-2}$Mpc.
This means that galaxies at recession velocities
$\simgt60{\,}000\,{\rm km}\,{\rm s}^{-1}$ should cease to show MOND effects.
Hence any galaxy or cluster at $z\simgt0.2$ should show no behaviour
normally taken to imply a dark halo!

As an example, let us consider the galaxy cluster MS\,1137,
at $z=0.783$ (Luppino \& Gioia~1995).
This is the second highest redshift cluster in the EMSS and has an x-ray flux
of $1.90\times 10^{37}h^{-2}\,{\rm W}$ in the $0.3$--$3.5\,$keV
band.  The x-ray temperature is $5.7^{+2.1}_{-1.1}$keV.
This cluster has been studied by Donahue et al.~(1999), who find that
the gas mass within $0.5h^{-1}$Mpc is
$1.2^{+0.2}_{-0.3}\times 10^{13}h^{-5/2}M_\odot$.
Using hydrostatic equilibrium and Newtonian mechanics (supposedly valid at
these high redshifts), they infer a total mass
$2.1_{-0.8}^{+1.5}\times 10^{14} h^{-1} M_\odot$ (within the same region).
This leads to a  gas fraction $f_{\rm gas}\,{=}\,0.06\pm 0.04 h^{-3/2}$, whereas
the MOND prediction would be $f_{\rm gas}\,{\equiv}\,1$.

There are certainly other examples of $z>0.2$ galaxies that do not appear
to behave any differently than their lower $z$ cousins.  Vogt et al.~(1997)
has at least one galaxy with a flat rotation curve at such a redshift.
Other work on the fundamental plane etc. in the most distant clusters would
be hard to explain if the outer regions of galaxies slowed down considerably
as MOND switched off.
Wilson et al.~(2001) studied statistical lensing around galaxies at
$z=0.1$--0.9.  Their results are consistent with what is expected from flat
rotation curves out to ${\sim}\,100\,$kpc,
with no sign of a redshift dependence.
This is hard to reconcile with a MOND picture in which the accelerations
go from the MOND to non-MOND regimes over this redshift interval.

However, Sanders (2000) suggested a different prescription for cosmological
accelerations in MOND.  The above objection
is avoided if one does {\it not\/} assume
an absolute meaning to Hubble accelerations relative to us.
Let us put aside the conceptual difficulties this leads to (that we already
discussed).  In this case, every observer's
Hubble acceleration still changes with time:
\begin{equation}
  {\ddot R\over R}  = H_0^2
  \left(\Omega_\Lambda - {1\over 2}\Omega_{\rm M}R^{-3}(t)\right).
\end{equation}
Thus at some time in the past ($z \sim 0.3$ for
a concordance cosmology and $z \sim 3$ for the parameters suggested
by McGaugh~2000), the
Hubble acceleration went through zero and hence cannot compensate for
difficulties arising from peculiar accelerations alone.  
At this time, all the MONDian
effects would come into force, for a period of time determined by
the size and other accelerations acting in the region considered.
Sanders (1998) discusses something similar this when considering a different
MONDian prescription.  At best we can say that when we consider accelerations
on cosmological scales the predictions of MOND are ambiguous.

\subsection{Gravitational Lensing in MOND}
Another entirely different constraint on
MOND comes from considerations of gravitational lensing.  The argument was
discussed in Walker (1994), but it is so nice that we repeat a modified
version here.  (Also see the recent papers by Mortlock \& Turner 2001a, 2001b).
Similar
arguments were also discussed by Edery \shortcite{Edery} and Bekenstein,
Milgrom \& Sanders \shortcite{BekMilSan}.

Kinney's (2000) argument about gravitational lensing of clusters of galaxies
shows that all modifications of gravity have to obey the equivalence
principle, in the sense that the dynamical mass must be the same as the mass
which couples to photons in lensing.
Once this is known, then we can treat the theory as a metric theory: the
objects involved know how to move independent of their mass, and thus the
`gravity' they feel can be written as a property of space-time independent
of the mass of the probe.
Walker (1994) argues that lensing effects at large distances can prove
disastrous for any theory in which potentials are not close to Newtonian
at the largest scales as well as locally.
His analysis shows that a
gravitational force which falls off like $\log(r)$ is inconsistent with the
mean convergence in the Universe being small.  With that much field at large
distances, the typical light ray gets strongly lensed, and objects are either
grossly magnified or demagnified.
Walker has estimated that the potential must become $1/r$ again beyond a radius
of about $10^{23}$m.
He notes that an analysis of any particular gravitational lens may put even
stronger constraints on the large-$r$ behaviour of gravity.

Whatever mimics the dark matter
has to couple to light the same way as to matter, or else you do not get the
masses of clusters correct (the factor of `2' in the bending of light is the
least of our worries!).
Then we can show that, generically, both the convergence and shear of a lens
scale as $a/r$ where $a(\vec{r})$ is the
bend angle at $\vec{r}$ and $\vec{r}$ is a 2D vector in the plane of the sky.
We are implicitly assuming here that $da/dr$ and $a/r$ are of the same order,
which should be true generally.
Now for a compact mass distribution in normal GR, the bend angle falls
off asymptotically as $1/r$, so our effect falls as $1/r^2$.
For a distribution of
sources, we need this fall-off for the total (integrated over $d^2r$) to be
a convergent function (remember the signs of the deflections are random so a
logarithmic divergence is really convergent).
If we have gravity falling off more slowly than $1/r$, then the bend angle
(which is just the line integral of $\nabla\Phi$ and so dimensionally goes
with the same power of $r$ as does $\Phi$)
also falls off more slowly than $1/r$ and
therefore the convergence (or shear) falls off more slowly than $1/r^2$ and
the integral over 2D space is divergent.

Hence, a theory with potentials falling off as $\log(r)$ will have huge
lensing optical depth, and rays will have chaotic paths through the distant
outer regions of galaxies.  This argument does not apply to the simplest
MOND picture, since that is not a metric theory.  It has been suggested that
there is some non-GR full theory, for MOND is just an approximation on the
scales of galaxies.  However, for even that to work, this full theory would
have to mimic GR closely on both small and large scales.  While clearly not
impossible, this does not seem very promising.  But, without anything like a
full theoretical framework, it is impossible to make definitive conclusions.

\section{The value of $\bmath{\lowercase{a}_0}$}
The value of the characteristic MOND acceleration scale $a_0$ is the same
order of magnitude as the dimensional acceleration $cH_0$.
It has been argued that there may be some fundamental reason behind this,
and that this provides a motivation for the notion that there is some deep
hidden truth underlying MOND.  It has also been suggested that this might
lead one to consider models in which $a_0$ varies over cosmological time
(e.g.~Milgrom 1994).  If the Universe is currently accelerating, then $H$
is smaller at high $z$, and so high redshift galaxies would be expected to
show less MOND effects if $a_0\propto H(z)$.  Taking the value today, if
$H_0\,{\simeq}\,70\,{\rm km}\,{\rm s}^{-1}{\rm Mpc}^{-1}$, then
$cH_0\,{\simeq}\,3.5a_0$, which is close, but not strikingly so.

{}From the point of view of the standard cosmological picture, this might be
thought of as merely a coincidence.  But as we shall argue, it is almost
inevitable that $a_0$ has this order of magnitude.  With some consideration of
dimensionless numbers and a little nod to the anthropic principle, the
coincidence is easily explained.

First notice that MOND was designed to fit galactic rotation curves, so
that $a_0\,{\sim}\,v^2/{\cal R}$ for the scale of a typical galaxy, ${\cal R}$.
Now notice that the dimensionless amplitude of potential perturbations
(one of the 6 fundamental cosmological numbers of Rees~1999) is given by
\begin{equation}
Q\sim{GM\over {\cal R} c^2}\sim{v^2\over c^2}\sim 10^{-5}
\end{equation}
over a wide range of cosmological scales.  Thus for MOND to fit galaxies we
need to have $a_0\,{\sim}\,Qc^2/{\cal R}$.  Then the MOND-cosmology coincidence,
$a_0\,{\sim}\,cH_0$ implies that ${\cal R}\,{\sim}\,Qc/H_0$, or in other words
\begin{equation}
{{\cal R}_{\rm galaxy}\over {\cal R}_{\rm Hubble}}\sim Q.
\end{equation}
To rephrase this, the mystery comes down to understanding why galaxies are
roughly $10^{-5}$ times the Hubble length.

Here we can appeal to arguments involving dimensionless numbers for setting
the sizes of stars and of galaxies (see e.g.~Padmanabhan~1996, \S1.17
and 1.19).  The consideration that a galaxy must be able to cool in a
gravitational collapse time implies that
\begin{equation}
{\cal R}_{\rm galaxy}\sim \alpha_{\rm G}^{-1} \alpha^3
 \left({m_{\rm p}\over m_{\rm e}}\right)^{1/2} {\hbar\over m_{\rm e} c},
\end{equation}
where $\alpha$ is the usual fine-structure constant and
$\alpha_{\rm G}\,{=}\,Gm_{\rm p}^2/\hbar c$
is the `fine-structure constant' for gravity.
A rather mild version of the anthropic principle suggests that the age of
the Universe today should not be too different from the characteristic ages
of stars, $t_\ast\,{\sim}\,\epsilon M_\ast c^2/L_\ast$, where $M_\ast$ and
$L_\ast$ are characteristic masses and luminosities, and $\epsilon$ is the
efficiency of nuclear energy generation.  $M_\ast$ can be estimated using
a well known argument that the temperature of a cloud of gas should be high
enough to allow nuclear fusion.  $L_\ast$ can then be estimated by assuming
that the photon opacity is dominated by Thomson scattering (for simplicity).
An estimate for the Hubble length is therefore given by $ct_\ast$, which is
\begin{equation}
{\cal R}_{\rm Hubble}\sim \alpha_{\rm G}^{-1} \alpha^{-1} \epsilon \eta^{-3/2}
 \left({m_{\rm p}\over m_{\rm e}}\right)^{1/2} {\hbar\over m_{\rm p} c},
\end{equation}
where $\eta$ is a correction factor for the nuclear tunnelling probability
(see Padmanabhan~1996 for details).

We might therefore expect
\begin{equation}
{{\cal R}_{\rm galaxy}\over {\cal R}_{\rm Hubble}}
 \sim \alpha^4 \epsilon^{-1} \eta^{3/2}
 \left({m_{\rm p}\over m_{\rm e}}\right).
\end{equation}
If we use typical values of $\epsilon\,{\sim}\,0.01$ and $\eta\,{\sim}\,0.1$,
we find that this ratio is indeed around $10^{-5}$, which is the value of
the dimensionless potential $Q$.
Within the standard cosmological framework this is just
a numerical coincidence, but a perfectly understandable one, without
any need to invoke MOND.

\section{Growth in High $\bmath{\Omega_{\rm B}}$ MOND models}

Now let us look in more detail at cosmological growth of perturbations and
the evolution of the power spectrum in a MOND picture.
Fig.~2 shows a particular example of evolution of a range of scales
in a MOND model, following
the arguments presented by Sanders (2000), which, as we have discussed,
are far from clear.
Let us analyse Sanders' (2000) results, since the final claim is that the
calculated MONDian  power spectrum has a general shape similar to the
$\Lambda$CDM one, and therefore can be considered in good agreement with the
data.

To re-cap, the idea is that in the early Universe the relevant
accelerations were larger, and hence MOND effects are negligible.  Whether this
is true or not depends, of course, on deciding which accelerations to choose,
as well as on having a consistent cosmological framework in which to discuss
cosmology at all.  For now we will assume that such difficulties will one
day be solved, and that the early Universe will behave
as in the conventional cosmology, so that we
can consider the approach Sanders (2000) has taken.
Here, the fluctuations would evolve according to 
standard Newtonian dynamics up to the point at which the Hubble acceleration
becomes small enough to trigger the MOND regime.
For scales between ${\sim}\,1$ and ${\sim}\,100\,$Mpc,
this happens in the matter dominated regime.
When MOND starts playing a role, the evolution of an overdensity follows 
a non--linear equation.
Note that, for smaller scales, this `switching redshift' would happen in the
radiation-dominated regime, and so more thorough calculations would have to
be done -- but this is a minor point.

One issue omitted from Sanders (2000) discussion is the following:
does the peculiar acceleration ever become big enough to re-establish Newtonian
dynamics?  Again we return to the whole ambiguous issue of the choice of
acceleration.  In any case we have numerically checked that under the
pseudo-Newtonian calculation of the MONDian evolution of a sphere
(assuming Sanders~2000 assumption about the correct acceleration to use),
structures on 70--$80\,$Mpc scales would become non--linear 
at $z\,{\simeq}\,1$ (see Fig.~2).

\begin{figure}
\begin{center}
\leavevmode
\epsfxsize=8cm \epsfbox{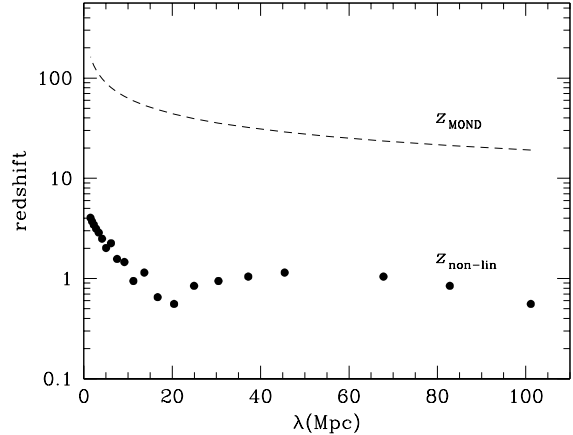}
\end{center}
\caption{Redshift at which a scale might enter the MOND regime
($z_{\rm MOND}$), and redshift at which it might become non-linear
($z_{\rm non-lin}$), according to the MOND-adapted
spherical top-hat ansatz suggested by Sanders (2000).  We have carried this
out for initial conditions for a particular $\Lambda$BDM cosmology (hence
the oscillations in $z_{\rm non-lin}$), and with $h=0.7$.
Although there are a number
of serious problems with this calculation, if taken at face value it
appears to show that in MOND even scales ${\sim}\,100\,$Mpc will have gone
non-linear today, in stark contradiction to many separate data-sets.}
\label{fig:znlmonznlmond}
\end{figure}

In general then, under this prescription we would expect structures to
collapse very much earlier than in the standard cosmology.  Star and/or
quasar formation would thus make the Universe highly ionized at high
redshift, erasing degree-scale CMB anisotropies, and possibly violating
the CMB spectrum $y$-distortion constraint.  In Fig.~1 then, the best
estimate for the CMB anisotropy spectrum may be one with significant
reionization optical depth (we show models with $\tau=0$, 0.5, 1 and 2).
Such a spectrum not only has a lower second
peak than expected, but no first peak to speak of either!

Moreover, for the matter perturbations, if we take the modified top-hat
calculations at face value, we find that scales as large as ${\sim}\,100\,$Mpc
may have already gone non-linear (Fig.~2) today.
This is entirely at odds with what we
know about large-scale power, where the perturbations have {\it low
amplitude\/} and appear to have close to {\it Gaussian statistics}.  How
this could be arranged  if the MOND perturbations have already gone non-linear
is hard to understand.   So the Sanders (2000) ansatz would seem to generate
massively {\it too much\/} power at large scales.

Now let us turn to the power spectrum shape.  The first question would be how
one even talks about a power spectrum in a framework which is inherently
non-linear.  This non-linearity means that $k$-modes are not independent
in MONDian cosmology, and hence it is not sufficient to evolve a range of
different $k$-modes in order to evolve the power spectrum.  Let us imagine,
however, that MONDian cosmology remains sufficiently linear until the
overdensities become themselves non-linear.
In that case the first thing to compute is simply the baryon-only model
power spectrum.
Fig.~1 shows that the power spectrum produced by a high $\Omega_{\rm B}$
model is extremely different from that produced by models with a significant
proportion of CDM, if we stick to Newtonian dynamics.
In addition to the total lack of small scale power, the
high $\Omega_{\rm B}$ model power spectra have enormous oscillations, going
out to very large scales.  Certainly there are plenty of data on relevant
scales (e.g.~Peacock \& Dodds~1996, Lin et al.~1996, Tadros \& Efstathiou
1996, Sutherland et al.~1999, Efstathiou \& Moody~2000,
Schuecker et al.~2001)
that show no indication of such dramatic oscillations.
While it is feasible (although not currently calculable in detail, since there
is no detailed theoretical framework in which to do so) that the effects of
MOND could induce new fluctuations on small scales, these would have to
contrive to fill in the oscillations in the power spectrum.  This seems a
very tall order, particularly on such large scales.
Since current data match the popular $\Lambda$CDM power spectrum rather well,
it is hard to see how any more detailed MOND-based calculation would rescue
these wildly different baryon-dominated power spectra.

Sanders (2000) suggests a recipe for computing a MOND power spectrum.
Since the MOND evolution equation is non-linear, 
he considers the evolution of an overdensity in real space,
rather than the more standard Fourier components of it.  The scale of the
spherical region in real space is identified with a wavenumber ${\sim}\,1/k$.
How precisely one interprets a set of spherical collapsing galaxies with
a random field of perturbations is unclear.
In computing the power spectrum, Sanders (2000) actually evolves 
$\Delta(k)=\sqrt{2\pi k^3 P(k)}$, which may not amount to the same result.
The final power spectrum is found reversing the same formula, after evolving
$\Delta(k)$ up to $z\,{=}\,0$.
While the lack of small scale power is partially rescued, it is still true 
that the imprint of the purely baryonic nature of the matter is manifest
through wild oscillations in the power spectrum.  In addition, although
Sanders (2000) finds that the resulting power spectrum qualitatively recovers
the small scale structure, the calculation is effectively modified from
linear theory, and hence does not contain the full non-linear growth.
Since fluctuations go non-linear much earlier in the MOND calculation
(see Fig.~2), the
final power spectrum would be considerably higher than claimed, and hence
in qualitatively poor agreement with data.

As an aside, there is an additional level of ambiguity in the way the
super-horizon scales are evaluated.  This implies an uncertainty on
the overall normalization of the power spectrum, and therefore on the
claimed value of $\sigma_8$.

Is there any way out for a model like MOND?  One possibility would be to
argue that complexities due to biasing or other non-linear effects could
make the measured power spectrum much smoother.
However, in the non-linear regime MOND$\to$GR and so this cannot save things;
it would be necessary to get rid of these oscillations in the linear regime.
To linear order you {\it cannot\/} couple Fourier modes, since differential
operators become linear in Fourier space.
Thus mode-coupling cannot wash out these oscillations.
Simple experiments we have done with bias also indicate it is difficult to
wash out these oscillations.  For any power-law scheme where
$\rho_{\rm gal}\propto \rho_{\rm mass}^{\rm B}$,
the oscillations survive in the galaxy power spectrum.

The only thing left then is to have growth rates which magically smooth the
spectrum out, i.e.~the peaks have to know to grow more slowly than the
troughs.  But the only new ingredient in the problem is $a_0$, otherwise
the standard Jeans growth analysis holds.  Since $a_0$ knows nothing about
the oscillation scale for the matter power spectrum, it is unclear how it
could smooth things out.  Or phrased another way, there can be at most a one
dimensional combination of parameters for which this is even conceivably true.
Finally, it is hard to see how initial conditions could have contrived to know
that these oscillations would be frozen in at $\sim1/3$eV by the recombination
of Hydrogen.

The bottom-line is that there is no legitimate way to carry out
calculations of the evolution of the power spectrum within MOND.  It is not
that the calculations are complicated, but that there is no framework in
which to carry out the calculations.  Hence any claims that MONDian cosmology
provides a good fit to CMB data or galaxy clustering are not currently
supportable.  As we have discussed, the opposite conclusion could just as
easily be reached.
The power on ${\sim}\,100\,$Mpc
scales is well in the linear regime and is comfortably fit by
{\sl COBE}-normalized $\Lambda$CDM models.  It seems unreasonable to appeal
to an argument that the calculations cannot currently by carried out, so maybe
MOND will eventually work well.  The simplest picture ($\Lambda$BDM) gives
disastrously low power on small scales, and oscillations on large scales, while
the modified spherical top-hat calculations indicate far too much power on
all relevant scales.

\section{Alternatives to MOND}

There have been several earlier ideas which share some similarities with
MOND.  We feel it is important to realise that there have been a great many
other suggestions for fitting flat rotation curves while avoiding dark matter.
Most of these were abandoned quite early on or never, in fact, taken as
serious alternatives in the first place.  

One of the earliest such proposals appears to have been by Finzi (1963),
who considered a law
of gravity which becomes stronger than Newton's law beyond some characteristic
length scale.
Tohline~(1983) discussed the possibility of the long-range force between
stars in galaxies becoming $1/r$.
Sanders~(1984) considered an effective anti-gravity force operating on scales
smaller than galaxies.
And Sanders~(1986) later considered a model which is like MOND but returns
to $1/r^2$ with a larger value of $G$ at scales much greater than that of
individual galaxies.
Bekenstein~(1988, see also Sanders~1989)
suggested the addition to GR of a complex scalar field whose phase couples
to ordinary matter, which gives force laws which behave similarly to MOND.
Meanwhile Talmadge et al.~(1988) set stringent limits on deviations from
$1/r^2$ force laws on solar system scales, and
McFarland (1990) placed similar constraints on $1/r^2$ deviations on
cluster scales.
Mannheim \& Kazanas (1989)
proposed a covariant theory which effectively has an extra
constant force at large radii.
Fahr (1990) suggested an additional inductive-type term in the gravitational
force due to mass currents.
Goldman et al.~(1992, see also Bertolami \& Garc{\' \i}a-Bellido 1996)
proposed a particular form of variable $G$ that might explain flat
rotation curves.
Battaner et al.~(1992) discussed the possibility
that magnetic fields could cause flat rotation curves without the need for
dark matter.
Gessner (1992) suggested that a large negative cosmological constant could
make rotation curves flat.
Eckhardt (1993) considered a two-parameter exponential potential.
Stubbs (1993) suggested the possibility of an exotic coupling between dark
matter particles and baryons, and set about constraining such a coupling.
Milgrom (1994) also suggested a model where dynamics might change below some
characteristic frequency rather than acceleration.
Soleng (1995) suggested that a point mass in a cloud of strings could give an
effectively $1/r$ force law.
Carlson \& Lowenstein (1996) considered the addition of a constant potential
term, which they claimed is motivated by conformal gravity.
Drummond (2001) suggested a bi-metric theory of gravity with a scale built in
at the size of galaxies, above which the effective value of $G$ is larger.

This is not a complete list, but shows some of the range of possibilities
that have been discussed.  In general, the concepts are phenomenological
and consider the solution of the dark matter `problem' to be confined to
explaining the haloes of individual galaxy-scale objects.
Most of these ideas must surely suffer from many of the shortcomings of MOND,
particularly
the large scale behaviour, which may be even more pathological in some cases.

Better motivated ideas of modifying gravity have also been attempted
(e.g.~Kinney \& Brisudova~2000), which also suffer from fatal flaws
(as they noted),
for example the problem with explaining gravitational lensing.
MOND needs to appeal to some as yet unknown process which reproduces the
usual `GR factor of 2' in order to explain lensing results (Qin et al.~1995)
for astrophysical objects like the sun, as well as for cosmological lensing .
For objects we would classically consider dark matter dominated, the situation
is much worse.  Consider, for example, clusters of galaxies, whose mass
estimated from dynamical arguments and gravitational lensing approximately
agree if $\Omega_{\rm CDM}\,{\simeq}\,10\times \Omega_{\rm B}$.
If light doesn't couple to gravity the same way as baryons, then you have a
factor of $\sim 10$ discrepancy.

Zhitnikov \& Nester (1994) presented a very elegant argument about the form
of reasonable extensions to GR which might explain flat rotation curves.
They assumed that such modifications are in the framework of metric theories
which also conform to several other rather weak assumptions.  The most
general metric that they found is more general than the usual
Parameterized Post-Newtonian formalism (see e.g.~Will 1993).
They found that constraints on the terms and symmetries within such a metric
(coming from the necessity to fit with solar system experiments,
gravitational deflection of light, etc.) make it seem rather improbable
that flat rotation curves can be explained away without the need for dark
matter.

MOND breaks several of the assumptions made by Zhitnikov \& Nester (1994).
However, the proponents of MOND suggest no other consistent framework with
which to replace the conventional picture.  This makes it currently
impossible to carry out definitive cosmological calculations in a MONDian
Universe.

\section{Conclusions}

Recently, there has been a renewed interest in the dark matter sector of the
standard cosmological theory.  In the absence of direct detection of a dark
matter candidate, this is understandable.  Some authors have even revived the
idea that a plausible explanation of galaxy rotation curves, one of the many
pieces of evidence for dark matter, lies not in the gravitational influence
of non-luminous matter but in a modification of our fundamental law of gravity.
Perhaps the longest lived such alternative is that of Milgrom, who proposed
the MOdified Newtonian Dynamics.

We have reviewed the extensive literature on MOND, including many studies which
show that MOND is both a {\it drastic\/} modification of our laws of motion and
gravity {\it and\/} that it fares poorly in explaining extant observations of
galactic structure.  We have included these since it is difficult to find a
comprehensive discussion of these issues published anywhere.  The focus of our
paper however is on the recent attempts to extend MOND to a theory of
cosmology.

We consider the numerous conceptual problems inherent in this approach, which
make MONDian cosmology much less than a theory.  We believe that the recent
claims of MONDian `predictions' or `cosmological models' are unsupportable.
Within the current MOND framework, the relevant calculations are fraught with
ambiguities and inconsistencies, making any quantitative calculations
impossible.

In our opinion, the entire premise of MONDian cosmology is at odds with the
modern view of the formation and evolution of galaxies.  The MOND calculations
are deeply rooted in the archaic models of monolithic galaxy formation, with
galaxies being unchanging, isolated, eternal objects whose centres define
special places in the universe about which accelerations can be measured.  It
is well nigh impossible to embed this theory within the modern context wherein
galaxies are interacting, dynamic objects, part of the evolving large-scale
structure of the universe.  MOND is at odds with the simplicity of hierarchical
clustering through gravitational instability which is a well-tested and
highly successful paradigm.  In this picture the hot early
universe was a simpler place, for which calculations can be done with
great precision.

\section*{ACKNOWLEDGMENTS}

EP and DS were supported by the Canadian Natural Sciences and Engineering
Research Council, JC and MW by the US National Science Foundation and MW
by a Sloan Fellowship.
We would like to thank St{\' e}phane Courteau for valuable discussions,
and Hilary Feldman for useful comments on the manuscript.
EP is grateful to Princeton University for hosting her during the preparation
of this work.  Alexander and Nicolo also contributed to the completion of
this work.

\end{document}